\documentclass[final]{aipproc}
\layoutstyle{6x9}
\usepackage{graphicx,amsmath,wrapfig,epsfig}
\begin{document}
\title{Recent progress \\ on nuclear parton distribution functions}
\author{M. Hirai}{
address={Department of Physics, Faculty of Science and Technology,
Tokyo University of Science \\ 
2641, Yamazaki, Noda, Chiba, 278-8510, Japan}}
\author{S. Kumano}{
  address={KEK Theory Center, Institute of Particle and Nuclear Studies, KEK \\
           and 
           Department of Particle and Nuclear Studies,
           Graduate University for Advanced Studies \\
           1-1, Ooho, Tsukuba, Ibaraki, 305-0801, Japan}}
\author{K. Saito}{
address={Department of Physics, Faculty of Science and Technology,
Tokyo University of Science \\ 
2641, Yamazaki, Noda, Chiba, 278-8510, Japan}}

\begin{abstract}
We report current status of global analyses on nuclear parton 
distribution functions (NPDFs). The optimum NPDFs are determined by
analyzing high-energy nuclear reaction data.
Due to limited experimental measurements, antiquark modifications
have large uncertainties at $x>0.2$ and
gluon modifications cannot be determined. 
A nuclear modification difference between $u$ and $d$ quark distributions 
could be an origin of the long-standing NuTeV $\sin^2 \theta_w$ anomaly. 
There is also an issue of nuclear modification differences between
the structure functions of charged-lepton and neutrino reactions.
Next, nuclear clustering effects are discussed in structure functions 
$F_2^A$ as a possible explanation for an anomalous result 
in the $^9$Be nucleus at the Thomas Jefferson National
Accelerator Facility (JLab).
Last, tensor-polarized quark and antiquark distribution
functions are extracted from HERMES data on the polarized structure
function $b_1$ of the deuteron, and they could be used
for testing theoretical models and for proposing future 
experiments, for example, the one at JLab. 
Such measurements could open a new field of spin physics 
in spin-one hadrons.
\end{abstract}

\maketitle

\section{Introduction}
\vspace{-0.10cm}

Parton distribution functions (PDFs) of the nucleon have been
investigated for a long time, and they are now established
except for extremely small-$x$ and large-$x$ regions.
A precise determination of the PDFs is very important for 
new discoveries at LHC (Large Hadron Collider). 
For heavy-ion collisions at LHC, nuclear parton 
distribution functions (NPDFs) should be also determined 
precisely because 10$-$20\% nuclear modifications exist
in medium-size nuclei and they are larger in heavy nuclei.
In addition to heavy-ion physics at LHC, an accurate determination 
of the NPDFs is valuable for other topics such as 
establishment of non-perturbative aspect of QCD by
comparisons with theoretical models of nuclear medium effects, 
neutrino-oscillation studies by nuclear corrections in $^{16}$O,
high-energy cosmic-ray interactions with air for studying GZK 
(Greisen-Zatsepin-Kuzmin) cutoff, and determination of 
fundamental constants like $\sin^2\theta_W$.
In contrast to the nucleonic PDFs, the NPDFs have not
been determined well, which adds uncertainties in investigating
the above-mentioned topics. In this report, we explain
the current status of the NPDF determination and our recent
studies on nuclear structure functions.

Nuclear modifications of the structure function $F_2$ have
been investigated in deep inelastic lepton scattering 
from nuclear targets \cite{sumemc}. Now, we have measurements 
from relatively small $x$ to large $x$ with a variety of nuclei
from the deuteron to heavy ones. Using these data together
with Drell-Yan measurements of nuclear targets, we determined
the NPDFs. RHIC (Relativistic Heavy Ion Collider) 
and neutrino reaction data could be used in addition. 
First, we introduce the current status of the NPDF determination
\cite{hkn,ds04,schienbein,eps,kp,fgs}
and a possible relation to the NuTeV $\sin^2 \theta_W$ anomaly
\cite{nutev-anomaly,sinth-npdf,sinth-cbt}.
We also discuss an issue of different nuclear modifications 
between charged-lepton and neutrino reactions.
Second, the structure function $F_2$ of $^9$Be is discussed 
for explaining an anomalous JLab data 
on the nuclear modification slope
$d(F_2^{Be}/F_2^D)/dx$ \cite{jlab-09}. 
Since $^9$Be is a nucleus which is usually described by 
clustering structure of two $\alpha$ nuclei with neutron clouds,
the data could indicate a clustering feature in the deep 
inelastic scattering (DIS) \cite{hksw-cluster}.
Third, new spin structure of the deuteron is investigated by 
analyzing HERMES data on the tensor-structure function $b_1$ 
\cite{hermes-b1,sk-b1-analysis}.
The tensor structure of nuclei and hadrons in the high-energy 
region is an unexplored topic, although it has been
investigated extensively at low energies in connection
with nuclear tensor force due to meson exchanges. It is interesting
to understand the tensor structure in terms of quark and gluon
degrees of freedom. Since $b_1$ vanishes if constituents
are in the $S$-wave, it could also be a good probe of 
quark-gluon dynamics including orbital angular momenta.

\section{Nuclear parton distribution functions}
\vspace{-0.10cm}

Nuclear modifications of the structure function $F_2$ are 
typically 10$-$20\% for medium size nuclei.
Therefore, it is convenient to express the NPDFs by the corresponding
nucleonic PDFs multiplied by nuclear modification factors $w_i$
($i=u_v,\ d_v,\ \bar q,\ g$):
\begin{equation}
f_i^A (x,Q_0^2) = w_i (x,A,Z) \, 
    \frac{1}{A} \left[ Z\,f_i^{p} (x,Q_0^2) 
                + N f_i^{n} (x,Q_0^2) \right] ,
\end{equation}
where $p$ and $n$ indicate the proton and neutron,
$A$, $Z$, and $N$ are mass, atomic, and neutron numbers,
respectively, and $Q_0^2$ is the initial $Q^2$ scale.
The isospin symmetry ($d^n  = u^p ,\;u^n  = d^p$)
is assumed in relating the distributions of the neutron to
the ones of the proton. It should be noted that the above 
parametrization cannot describe the NPDFs at $x>1$ because 
the distributions vanish in the nucleon, whereas the kinematical
range is $0<x<A$ in a nucleus.

There are a few groups which have been investigating the optimum
NPDFs by global analyses of world data on high-energy nuclear
reactions \cite{hkn,ds04,schienbein,eps,kp,fgs}.
The $x$-dependent functional forms are different
depending on the analysis groups. For example, the HKNS 
parametrization uses \cite{hkn}
\begin{equation}
w_i (x,A,Z) = 1 + \left( {1 - \frac{1}{{A^{\alpha} }}} \right)
\frac{{a_i  + b_i x + c_i x^2  + d_i x^3 }}{{(1 - x)^{\beta} }} ,
\\
\end{equation} 
at $Q_0^2=1 \ {\rm GeV}^2$. Here, $\alpha$, $\beta$, $a_i$, $b_i$,
$c_i$, and $d_i$ are parameters to be
determined by a $\chi^2$ analysis.
Other groups' functional forms are listed in Ref. \cite{nuint09-npdf}.
In the parametrization of de Florian and Sassot \cite{ds04}, 
the NPDFs are expressed by convolution of the nucleonic
PDFs with a modification function $w(x)$ and a small initial
$Q^2$ scale (0.4 GeV$^2$) is used for utilizing the GRV-type PDFs.
In the analysis of Eskola, Paukkunen, and Salgado (EPS) \cite{eps},
the $x$ region is divided into three: shadowing, anti-shadowing,
and EMC \& Fermi-motion regions, and different parameters are
assigned to the three functions in each NPDF.
Schienbein {\it et al.}'s analysis used a functional form
of polynomials of $x$ combined with exponential functions
\cite{schienbein}. Used data sets are similar; however, the 
EPS \cite{eps} included some of the RHIC data and Schienbein {\it et al.}
\cite{schienbein} also used NuTeV neutrino data.
There are other works by Kulagin and Petti (KP) \cite{kp}
and Frankfurt, Guzey, and Strikman (FGS) \cite{fgs}.
The KP method is to determine off-shell structure
functions of the nucleon by calculating other nuclear
modifications in conventional models with binding, Fermi motion, 
and multiple scattering. 
The FGS method is mainly intended to describe the small-$x$ 
shadowing part by the multiple scattering.

   \begin{center}
\begin{figure}[h]
   \vspace{-0.30cm}
       \epsfig{file=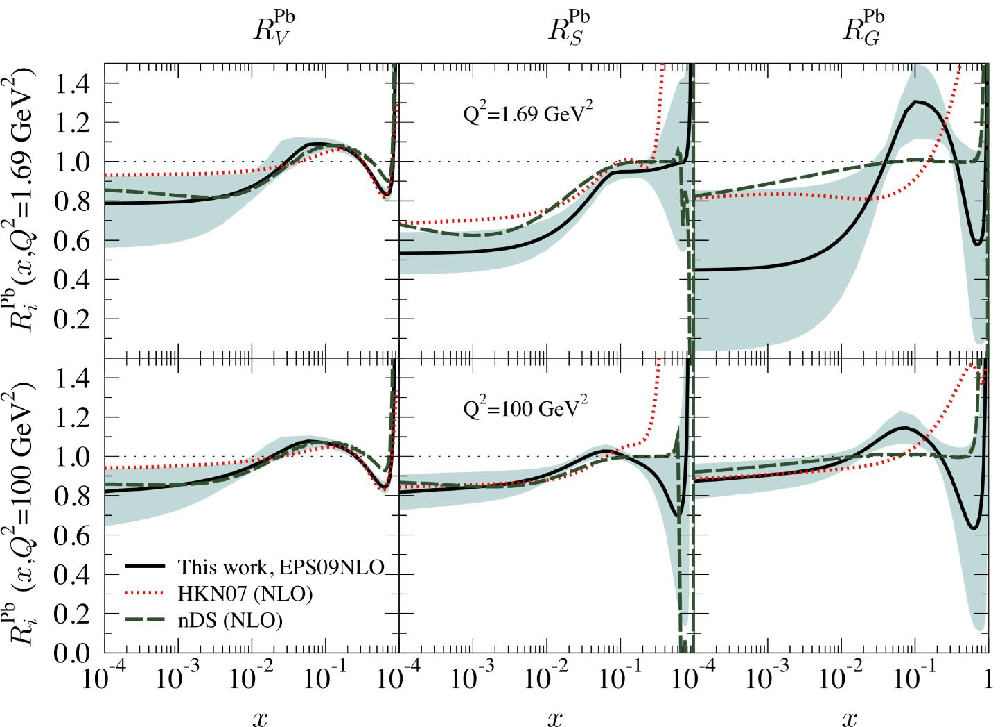,width=0.65\textwidth} 
\end{figure}
\noindent
   \begin{minipage}[c]{10.0cm}
   \setlength{\baselineskip}{10pt} 
   {\footnotesize 
   {\bf FIGURE 1.} Nuclear modifications of the PDFs are shown
   for the lead nucleus at $Q^2$=1.69 and 100 GeV$^2$ \cite{eps}.
   The functions $R_V^{Pb}$, $R_S^{Pb}$, and $R_G^{Pb}$ indicate
   the valence-quark, sea-quark, and gluon modifications,
   respectively. Three parametrization results (EPS09, HKN07, and nDS)
   are shown with uncertainty bands.
}
   \end{minipage}
   \end{center}
\vspace{0.1cm}

Nuclear modifications of the PDFs are determined by $\chi^2$
analyses of world data on the structure-function $F_2$ 
ratios $F_2^A / F_2^{A'}$, Drell-Yan cross-section ratios 
$\sigma_{DY}^{pA} / \sigma_{DY}^{pA'}$, RHIC $dA$ data, 
and neutrino-DIS structure functions. Typical results are shown 
in Fig. 1 for nuclear modifications of the lead nucleus 
at $Q^2$=1.69 and 100 GeV$^2$.
Three different parametrizations are shown together with 
uncertainty bands. Although the three groups used 
different parametrizations for the $x$ dependence and 
slightly different data sets, the determined distributions
roughly agree with each other within the uncertainty bands.
There are large uncertainties in the antiquark distributions 
at medium and large $x$ and in the gluon distribution 
at whole $x$. They should be determined by future experimental
measurements.

The studies of nuclear modifications could affect 
the NuTeV anomaly on the weak-mixing angle. 
The NuTeV anomaly has been a long-standing issue that
the mixing angle measured by the NuTeV neutrino DIS, 
$\sin^2 \theta_W = 0.2277 \pm 0.0013 \, \text{(stat)} 
                         \pm 0.0009 \, \text{(syst)}$,
is different from other measurements, 
$\sin^2 \theta_W = 0.2227 \pm 0.0004$ \cite{nutev-anomaly}.
In extracting the mixing angle from neutrino DIS,
there is a Paschos-Wolfenstein (PW) relation
for the isoscalar nucleon. There are various corrections
to use it for the iron target in the NuTeV experiment:
\begin{equation}
\frac{  \sigma_{NC}^{\nu A}  - \sigma_{NC}^{\bar\nu A} }
              {   \sigma_{CC}^{\nu A}  - \sigma_{CC}^{\bar\nu A} }
        =  \frac{1}{2} - \sin^2 \theta_W 
        +\text{($Z \ne N$)}
        +\text{($w_{u_v} \ne w_{d_v}$)}
        +\text{($u^p \ne d^n$)}
        +\text{($s \ne \bar s$)}
        +\text{($c \ne \bar c$)}
\, ,
\label{eqn:pw}
\end{equation}
where $\sigma_{CC}^{\nu A}$ and $\sigma_{NC}^{\nu A}$ are
charged-current (CC) and neutral-current (NC) cross sections,
respectively, in neutrino-nucleus DIS.
The parentheses $(\cdot\cdot\cdot)$ indicate various corrections
to the PW relation. The obvious non-isoscalar correction ($Z \ne N$)
is taken into account in the NuTeV analysis, so that it should
not be the origin of the anomaly.

First, since the NuTeV target is the iron nucleus, the anomaly
could come from nuclear modification difference between $u_v$
and $d_v$. In particular, baryon-number and charge conservations
indicate that the modifications should be different between 
$u_v$ and $d_v$ \cite{sinth-npdf}. However, the current data 
are not enough to probe such a small effect by a global NPDF 
analysis \cite{sinth-npdf}. According to Clo\"et, Bentz, and Thomas,
the nuclear modification is larger for $u$ quark than the one
for $d$ due to vector mean field which is adjusted to
explain the symmetry energy of a nucleus \cite{sinth-cbt}.
Future pion-induced Drell-Yan measurements such as
by the COMPASS collaboration could be used for finding 
the difference \cite{udv-npdf-dy}.
Second, QED effects on the PDFs, namely $u^p \ne d^n$ and 
$d^p \ne u^n$ could be the origin of the anomaly. The symmetry
relations between the proton and neutron PDFs ($u^p = d^n$, $d^p = u^n$)
are usually assumed in all the global analyses by considering 
that the QED effects are less than a few percent 
($\sim\alpha/\alpha_s$). 
A possible QCD effect was investigated in Ref. \cite{mrst-qed} 
by somewhat assuming a photon distribution at the initial 
scale $Q^2$=1 GeV$^2$. The result indicated that the anomaly 
could come from such an effect although a precise determination
of the QED effect is not possible at this stage.
Third, the strange and charm asymmetries ($s\ne \bar s$, $c\ne \bar c$)
could be origins. However, such effects cannot be determined from
experimental measurements at this stage in the sense that 
the distribution $s(x)-\bar s(x)$ still has a large uncertainty 
band \cite{s-sbar}.

\begin{wrapfigure}{r}{0.52\textwidth}
   \vspace{-0.1cm} \hspace{0.5cm}
       \epsfig{file=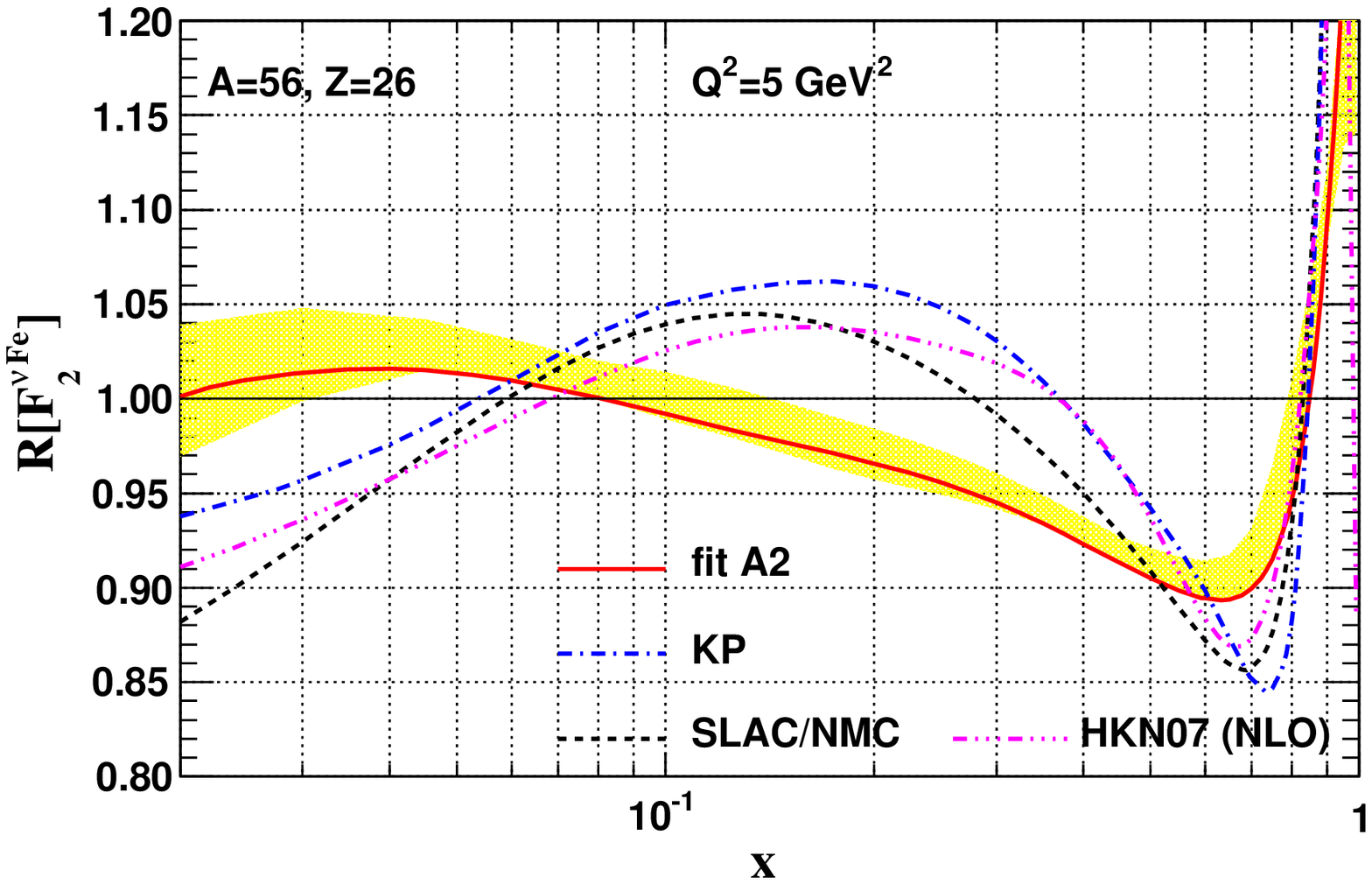,width=0.50\textwidth} \\
   \vspace{-0.6cm}
       \begin{minipage}[c]{0.4cm}
       \ \ 
       \end{minipage}
       \begin{minipage}[c]{7.0cm}
       \setlength{\baselineskip}{10pt} 
\vspace{-0.35cm}
{\footnotesize {\bf FIGURE 2.} 
   The solid curve (fit A2) indicates the nuclear modification
   of $F_2$(iron) in neutrino DIS \cite{schienbein}. 
   For comparison, two parametrization results (KP and HKN07)
   are shown together with the modification in charged-lepton
   DIS (SLAN/NMC).
}
       \end{minipage}
   \vspace{+0.6cm}
\end{wrapfigure}

Another issue of the current NPDFs is the modification differences
between charged-lepton and neutrino reactions \cite{schienbein}. 
The nuclear modifications of $F_2$ are well established in 
charged-lepton reactions, and the optimum NPDFs are extracted 
mainly from their data. However, the modifications are very different 
in neutrino DIS as shown by the solid curve in Fig. 2 \cite{schienbein}, 
whereas the other curves are the modifications 
suggested by charged-lepton DIS.
The medium-$x$ ($= 0.5 \sim 0.8$) modifications are different
and the antishadowing phenomenon does not appear at $x = 0.1 \sim 0.2$
in the neutrino DIS. 
Since neutrino data are partially used for extracting current
``nucleonic" PDFs, this issue should be solved for accurate
determinations of not only the NPDFs but also the PDFs 
of the nucleon. 

\vspace{-0.15cm}
\section{Clustering effects in nuclear structure functions}
\vspace{-0.15cm}

It is known in nuclear-structure studies
that some light nuclei exhibit phenomena of nuclear 
clustering. For example, the $^8$Be nucleus is theoretically
described by two $\alpha$-like clusters according to a Monte Carlo 
calculation \cite{mc-8be-2000} rather than the usual monotonic 
density distribution estimated by a shell model. This fact 
led us to wonder whether such clustering structure could 
appear in high-energy reactions, for example, in
the structure functions $F_2^A$.
Although simple multiquark clusters were investigated
at the early stage of EMC-effect studies, clustering
aspects of nucleons were not investigated. Recently, such 
a signature could be found by a JLab experiment \cite{jlab-09}
according to the theoretical work of Ref. \cite{hksw-cluster}. 
We explain this theoretical approach on the clustering effects.

First, we estimate effects of nuclear clustering by using
a conventional convolution approach. 
A nuclear structure function $F_2^A$
is given by the nucleonic one convoluted with 
a nucleon momentum distribution in the nucleus $f(y)$:
\cite{sumemc,ek03}
\begin{align}
\! \! \! \! \!
F_{2}^A (x, Q^2) & = \int_x^A dy \, f(y) \, F_{2}^N (x/y, Q^2) , \ \ 
f(y)  =  \frac{1}{A} \int d^3 p_N
     \, y \, \delta \left( y - \frac{p_N \cdot q}{M_N \nu} \right) 
     | \phi (\vec p_N) |^2 ,
\label{eqn:w-convolution}
\end{align}
where $F_2^N$ is the structure function for the nucleon, 
and $y$ is the momentum fraction 
$ y   =  M_A \, p_N \cdot q /(M_N \, p_A \cdot q)
  \simeq A \, p_N^+ /(p_A^+) $
with a light-cone momentum $p^+$.
If the momentum density $| \phi (\vec p_N) |^2$ is obtained,
$F_2^A$ can be calculated by this equation.

\begin{figure}[b]
\begin{minipage}[c]{\textwidth}
\vspace{-0.3cm}
   \includegraphics[width=0.47\textwidth]{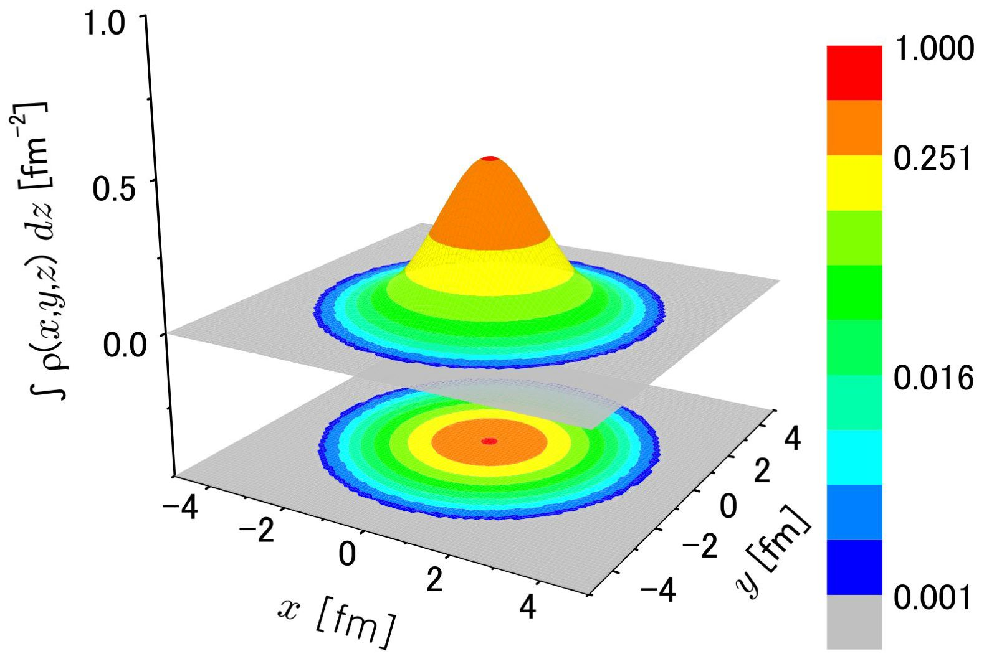}
   \hspace{0.4cm}
   \includegraphics[width=0.47\textwidth]{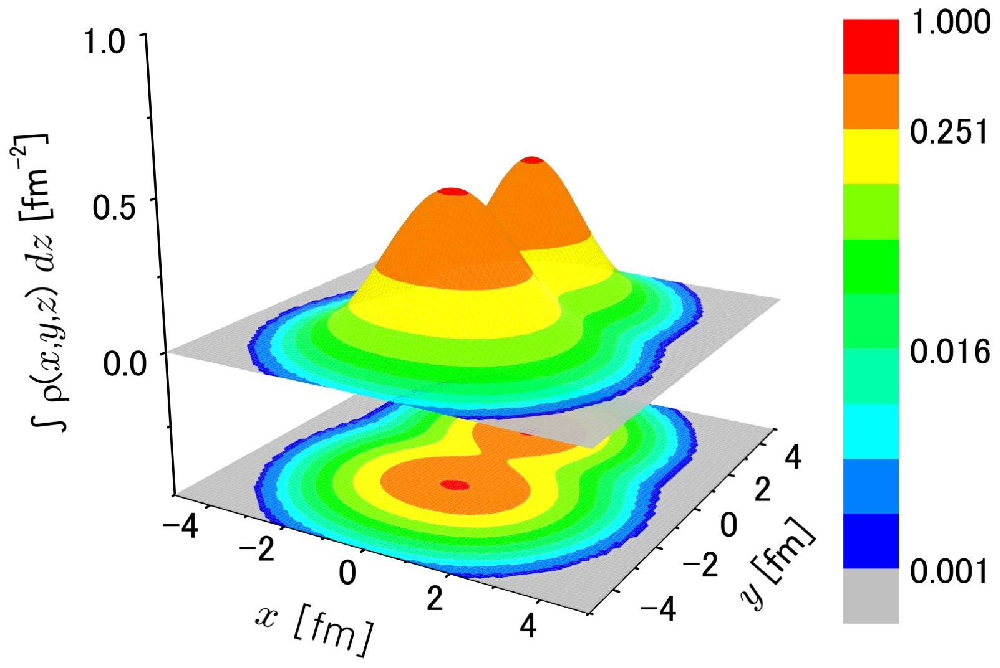}
 $ \ $
\vspace{+0.3cm}
\noindent
\begin{minipage}[c]{0.47\textwidth}
\setlength{\baselineskip}{10pt}
\vspace{0.3cm}
{\footnotesize 
 {\bf FIGURE 3.} 
 Two-dimensional ($x,\, y$) coordinate-space density of 
 $^4$He by the AMD \cite{hksw-cluster}.
 The density is shown by taking an integral over 
         the coordinate $z$: $\int dz \rho(x,y,z)$.
 The shell-model density is almost the same as
 the above distribution in $^4$He.
}
\end{minipage}
\vspace{-1.60cm}\hspace{0.50cm}
\begin{minipage}[c]{0.47\textwidth}
\setlength{\baselineskip}{10pt}
\vspace{0.3cm}
{\footnotesize 
 {\bf FIGURE 4.} 
 Two-dimensional ($x,\, y$) coordinate-space density of 
 $^9$Be by the AMD \cite{hksw-cluster}. 
 The notations are the same as the one in Fig. 3.
 The shell-model density is a monotonic distribution, which is 
 different from the above distribution.
 }
\end{minipage}
\end{minipage}
\end{figure}

Two theoretical models are used for calculating the nuclear densities.
In order to investigate a possible clustering effect, 
an antisymmetrized molecular dynamics (AMD) is used and
a simple shell model is also employed to compare with 
AMD calculations. The AMD is a variational approach 
for describing nuclei. Its advantage is that there is 
no a priori assumption on nuclear structure whether
it is a shell or cluster-like configuration.
A nuclear wave function is given by the Slater determinant
of single-particle wave packets:
$
\left | \Phi (\vec r_1, \vec r_2, \cdot\cdot\cdot, \vec r_A ) \right >
       =  ( 1 / \sqrt{A!} )
            \text{det} [ \varphi_1 (\vec r_1), \varphi_2 (\vec r_2), 
                                 \cdot\cdot\cdot, \varphi_A (\vec r_A) ] 
$
where $\varphi_i (\vec r_j)$ is the single-particle wave function
given by
$\varphi_i (\vec r_j) = \phi_i (\vec r_j) \, \chi_i \, \tau_i$
with spin and isospin states $\chi_i$ and $\tau_i$. 
The function $\phi_i (\vec r_j)$ is 
the space part expressed as 
$
\phi_i (\vec r_j) = ( 2 \nu / \pi )^{3/4}
        \exp  [ - \nu 
           ( \vec r_j - \vec Z_i / \sqrt{\nu} ) ^2  ] ,
$
where $\nu$ and $\vec Z_i$ are variational parameters.
Simple effective $NN$ interactions are used for describing gross
properties of nuclei. The variational parameters are then determined
by minimizing the system energy with a frictional-cooling method.
Obtained space densities are shown in Figs. 3 and 4 for $^4$He 
and $^9$Be.
Figure 3 indicates that the $^4$He density is a monotonic distribution
as typically suggested by shell models. However, it is obvious 
from Fig. 4 that the $^9$Be distribution is totally different. There are
two density peaks which roughly correspond to two $\alpha$ clusters
with surrounding neutron clouds. It is interesting to investigate
that such a nuclear clustering could be found in high-energy
experiments, for example, in nuclear structure functions because
the anomalous EMC effect was indicated by the JLab experiment
for $^9$Be \cite{jlab-09}.

In order to show the clustering effects, a simple shell-model
is used for calculating nuclear wave functions in comparison
with the AMD results. 
We use a harmonic-oscillator model with the potential 
$M_N \omega^2 r^2/2$.
Its wave function is given by
$
\psi_{n \ell m} 
         = R_{n \ell} (r) Y_{\ell m} (\theta, \phi) , \ \ \ 
R_{n \ell} (r) 
         = N_{n \ell}
                  r^\ell e^{-\frac{1}{2}\kappa^2 r^2} 
                  L_{n-1}^{\ell+1/2} (\kappa^2 r^2) ,
$
where $Y_{\ell m} (\theta, \phi)$ is the spherical harmonics,
$L_{n-1}^{\ell+1/2} (x)$ is the Laguerre polynomial, $\kappa$ 
is defined by $\kappa = \sqrt{M_N \omega}$,
and $N_{n \ell}$ is the normalization constant.
The shell-model density of $^9$Be is a monotonic distribution,
which is totally different from the AMD one in Fig. 4. 
However, if an angular average is taken for the density,
the difference between the AMD and shell model is not
very large \cite{hksw-cluster}.
Taking the Fourier transformation to the momentum space,
we show both densities in Fig. 5 for $^4$He and $^9$Be. 
Although the AMD and shell-model densities are almost the same
in the $^4$He nucleus, they are different in $^9$Be.
The AMD distribution is shifted toward the high-momentum
region, which is caused by the nuclear clustering. 
Since nucleons are confined mainly in two small space 
regions (clusters), the high-momentum components are created.
It is interesting to find such a difference due to the nuclear 
clustering. We then estimate its effect on nuclear structure 
functions $F_2^A$ by using Eq.(\ref{eqn:w-convolution}).

\begin{figure}[b]
\begin{minipage}[c]{\textwidth}
   \includegraphics[width=0.47\textwidth]{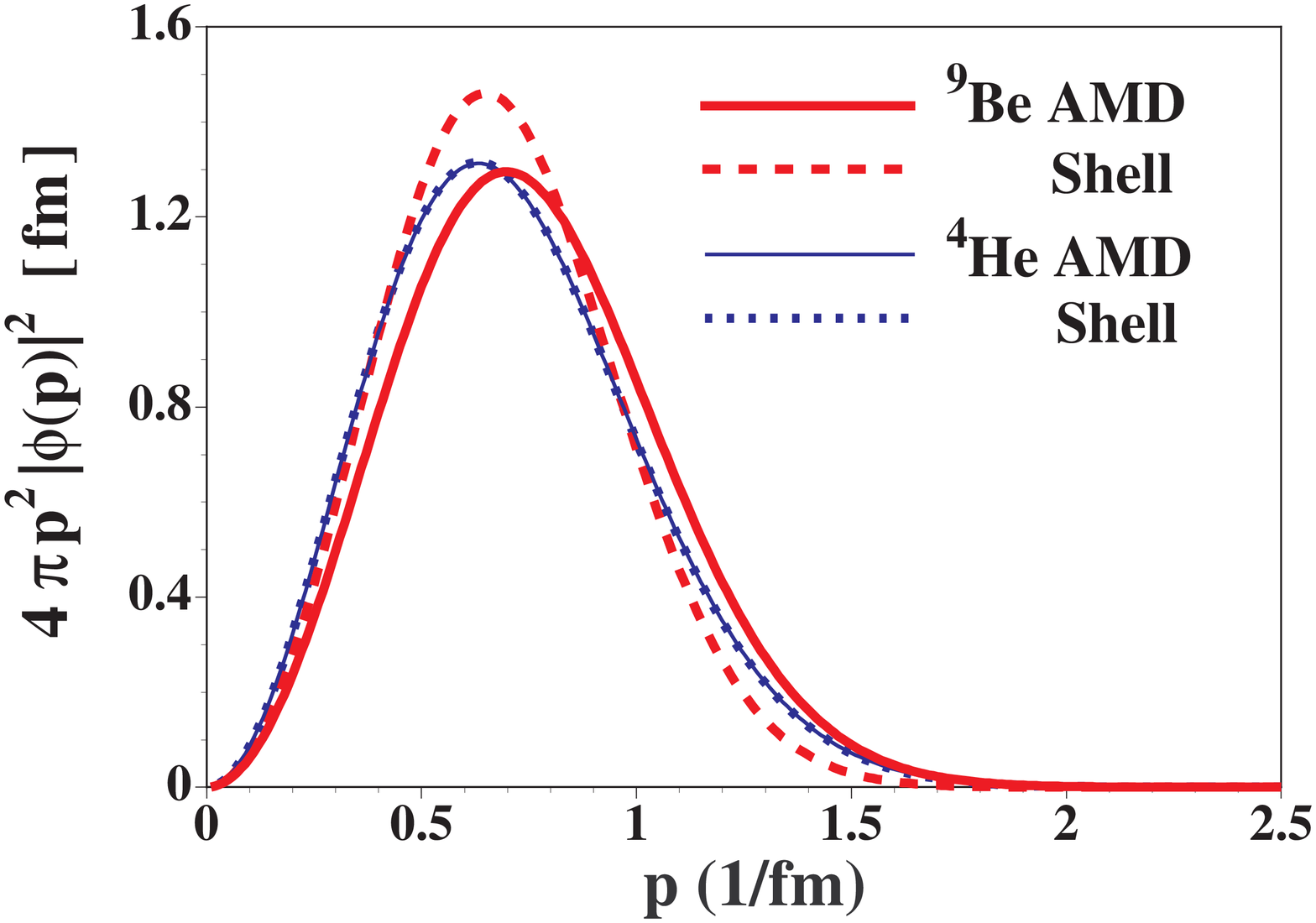}
   \hspace{0.6cm}
   \includegraphics[width=0.46\textwidth]{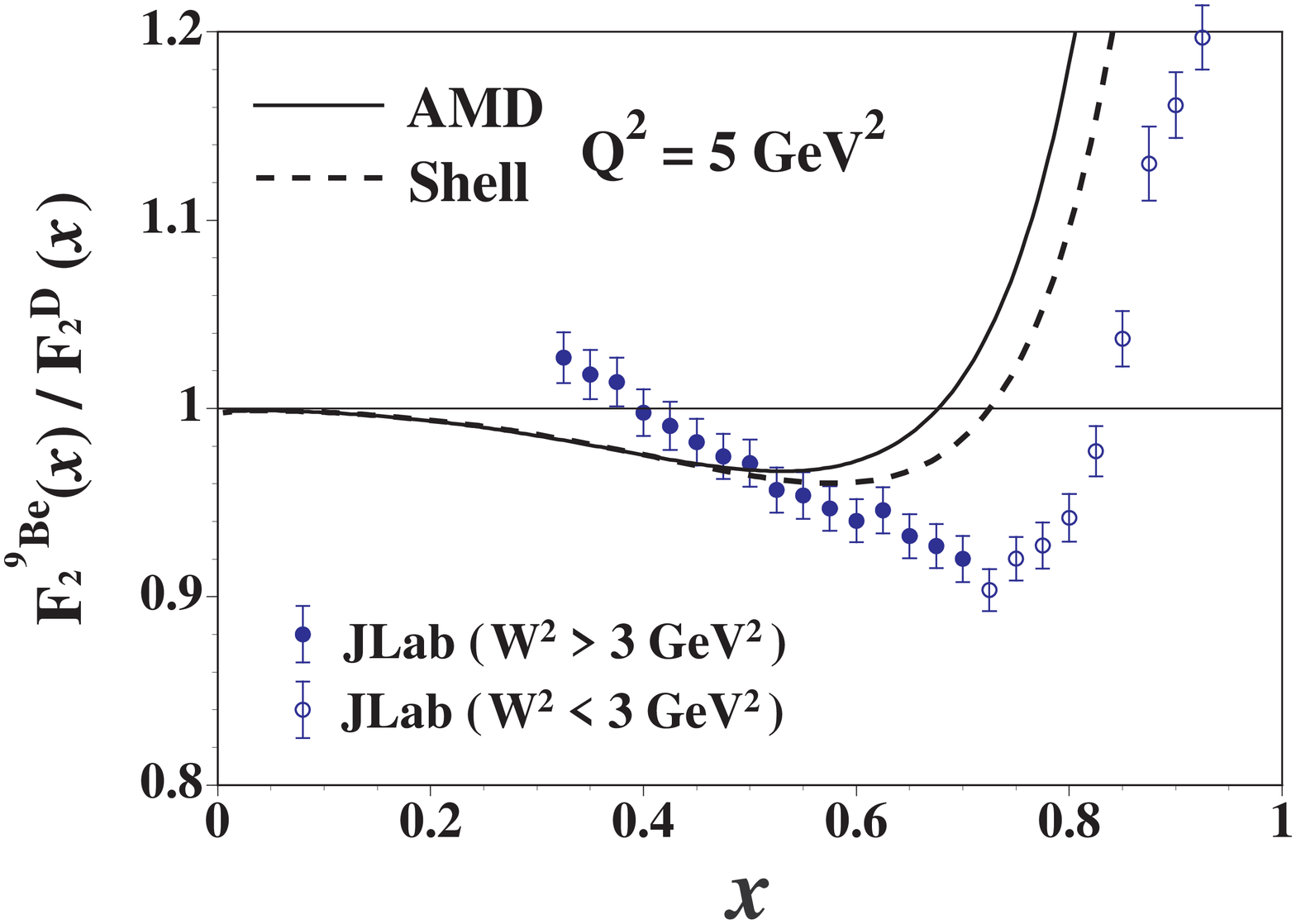}
  $ \ \ $
\label{fig:b1-motivation}
\vspace{-0.1cm}
\noindent
\begin{minipage}[c]{0.47\textwidth}
\setlength{\baselineskip}{10pt}
\vspace{0.3cm}
{\footnotesize 
 {\bf FIGURE 5.} Nucleon-momentum distributions in $^4$He and $^9$Be
                 by the shell and AMD models \cite{hksw-cluster}.
                 They are obtained by taking Fourier transformations
                 of the coordinate-space densities.} 
\end{minipage}                 
\hspace{0.7cm}
\begin{minipage}[c]{0.46\textwidth}
\setlength{\baselineskip}{10pt}
\vspace{0.3cm}
{\footnotesize 
 {\bf FIGURE 6.} 
 Nuclear modifications of $F_2$ for $^9$Be \cite{hksw-cluster}. 
 The dashed and solid curves indicate
 the shell-model and AMD results at $Q^2$=5 GeV$^2$.
 JLab measurements are shown for comparison.}
\end{minipage}
\end{minipage}
\end{figure}
\vspace{0.0cm}

The EMC ratio ($F_2^A/F_2^D$) is shown for the $^9$Be nucleus
in Fig. 6 together with the JLab data. Since effects of
short-range correlations and internal nucleon modifications
are not included in our estimate, the obtained curves do
not fully agree with the data. Here, it is our intention to 
show the clustering effect in the basic convolution formalism.
Considering that the differences between curves are of the order 
of the experimental errors and there are some theoretical
ambiguities, for example, from the correlations and nucleon 
modifications if they are included in our estimate,
it may not seem easy to find a clear clustering signature. 

\begin{figure}[t]
\begin{minipage}[c]{\textwidth}
   \includegraphics[width=0.46\textwidth]{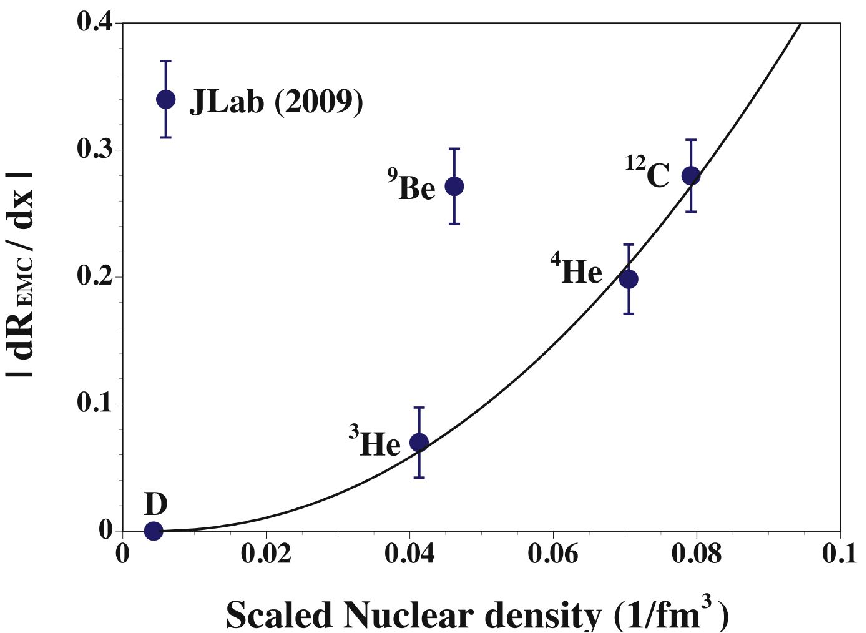}
   \hspace{0.7cm}
   \includegraphics[width=0.46\textwidth]{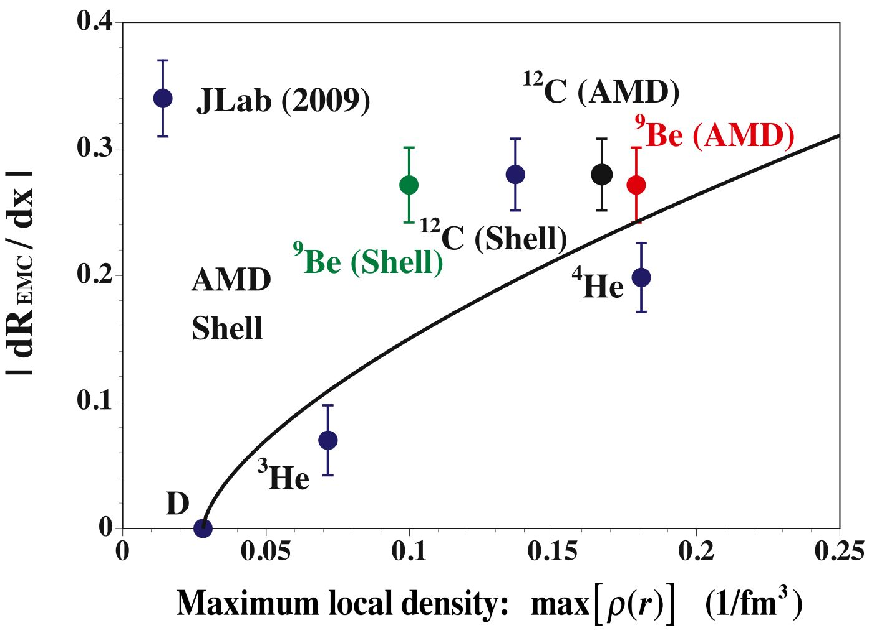}
  $ \ \ $
\label{fig:b1-motivation}
\noindent
\begin{minipage}[c]{0.46\textwidth}
\setlength{\baselineskip}{10pt}
\vspace{0.4cm}
{\footnotesize 
 {\bf FIGURE 7.} Nuclear modification slopes measured at JLab
 are shown by taking the scaled nuclear density 
 as the abscissa \cite{jlab-09}.
 The curve indicates a fit to the data except for $^9$Be.} 
\end{minipage}
\vspace{-0.00cm}\hspace{0.6cm}
\begin{minipage}[c]{0.47\textwidth}
\setlength{\baselineskip}{10pt}
\vspace{0.4cm}
{\footnotesize 
 {\bf FIGURE 8.} Nuclear modification slopes are shown
                 by taking the maximum local densities 
                 in the shell and AMD models
                 as the abscissa \cite{hksw-cluster}.
       The curve indicates a fit to the data points of the shell model
       except for $^9$Be.}
\end{minipage}
\end{minipage}
\end{figure}
\vspace{0.0cm}

The Jlab experimental group obtained the nuclear modification 
slope $dR/dx$ for $R=F_2^A/F_2^D$ by approximating their data 
by straight lines at $0.35<x<0.7$ \cite{jlab-09}. They are shown 
in Fig. 7 as a function of the scaled nuclear density. 
All the data are along the smooth curve, whereas the $^9$Be 
slope cannot be explained by such smooth density dependence.
However, if the maximum local density is taken as the abscissa,
the situation is different as shown in Fig. 8. The curve 
is plotted to interpolate the shell-model slopes except
for $^9$Be. If the maximum density is calculated by the shell model,
the modification slope $|dR/dx|$ is too large to be expected
from the $^9$Be density, whereas it is almost on the curve
if the maximum density is calculated in the AMD with 
the clustering structure. The ``anomalous'' JLab result on the $^9$Be
could be explained by the cluster structure in the nucleus.
We also notice a small difference between 
the AMD and shell model in $^{12}$C of Fig. 8, and
it is caused by the mixing of a cluster-like configuration 
in the AMD result of $^{12}$C.

We would like to clarify our viewpoint because it may be confusing
to the reader why the clustering effects appear in $dR/dx$
of Fig. 8, whereas the effects are small in Fig. 6. 
The nuclear structure functions consist of the mean conventional 
part and the remaining one depending on the maximum local density.
The first part is described by the usual convolution calculation
with the spectral function given by the averaged nuclear density 
distribution, and thus the inhomogeneity of the nuclear density
is washed out. The remaining second part is associated with
the inhomogeneity of the nuclear density, before taking 
the average of nuclear wave function, given by the nuclear 
cluster structure. 
Our results imply that the physics mechanism, associated 
with the high densities created by the clusters in $^9$Be,
could be the origin for the anomalous 
nuclear-modification slopes $dR/dx$ of $^9$Be.

\vspace{-0.1cm}
\section{Tensor structure by quark degrees of freedom}
\vspace{-0.2cm}

Tensor structure has been investigated in nuclei by hadron degrees
of freedom, namely by nucleons and mesons. For example, the $D$-state
admixture due to tensor force in nucleon-nucleon interactions is
the reason for the finite quadrupole moment in the deuteron. However,
the time has come to investigate the tensor structure by quark
degrees of freedom. It can be studied by measuring the tensor
structure function $b_1$ in deep inelastic lepton-deuteron 
scattering as illustrated in Fig. 9.

\vspace{0.3cm}
\begin{figure}[h]
   \includegraphics[width=0.85\textwidth]{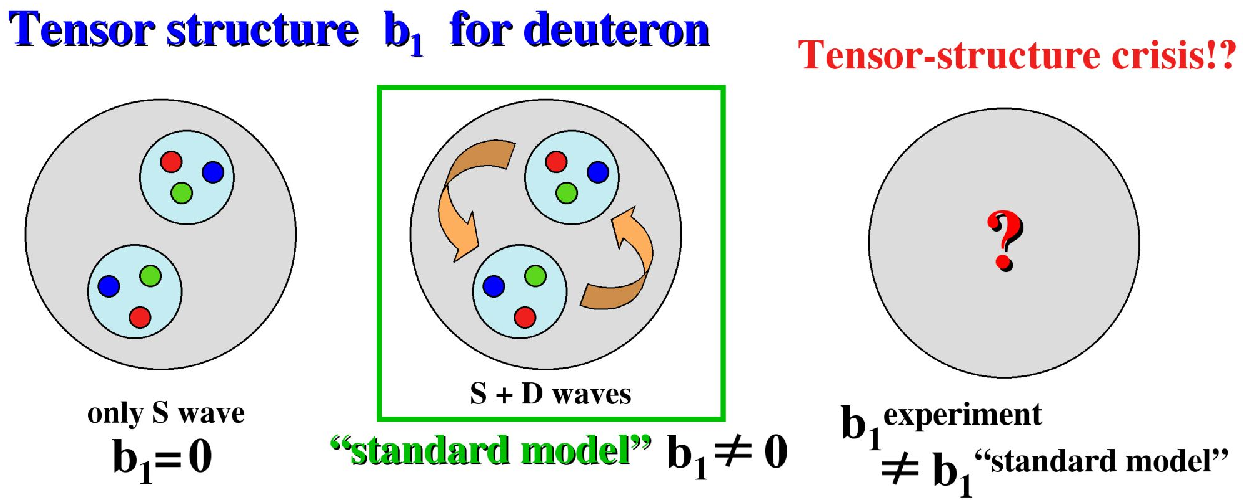}
\label{fig:b1-motivation}
\end{figure}
\vspace{-0.1cm}
\hspace{0.6cm}
\begin{minipage}[c]{0.85\textwidth}
\setlength{\baselineskip}{10pt}
\vspace{0.2cm}
{\footnotesize {\bf FIGURE 9.} Illustration of tensor structure
 in the deuteron. The structure function $b_1$ vanishes if 
 constituents are in the S-wave. A finite $b_1$ can be obtained
 in the conventional deuteron model with the D-state admixture,
 and it may be considered as the ``standard model". 
 Nucleon-spin studies imply that such a conventional picture
 could fail. Then, a new field of spin physics will be explored
 by future experimental measurements.
 }
\end{minipage}
\vspace{0.6cm}

It could be predicted in a conventional nuclear-physics approach
with the usual convolution description by PDFs 
convoluted with nucleon-momentum distributions
as illustrated in the middle of Fig. 9.
However, our experience of nucleon-spin puzzle suggests that
such a traditional model would not work in spin and orbital 
angular momenta. There is a good possibility that interesting
new field of physics will be created by measuring the tensor
structure functions.

The structure function $b_1$ is expressed by 
tensor-polarized distributions $\delta_T q$ as
\begin{equation}
b_1 (x) = \frac{1}{2} \sum_i e_i^2 
   \left[ \delta_T q_i(x) 
        + \delta_T \bar q_i(x) \right ] , \ \ \ 
\delta_T q_i (x) = q^0_i (x)
    -\frac{q^{+1}_i (x) +q^{-1}_i (x)}{2}  , 
\end{equation}
where $i$ indicates the flavor of a quark, $e_i$ is the charge
of the quark, $q_i^\lambda$ indicates an unpolarized-quark distribution
in the hadron spin state $\lambda$, and the distributions
are defined by the ones per nucleon.
For analyzing HERMES data on $b_1$, the distributions are
expressed as
$\delta_T q_{iv}^D (x) = \delta_T w(x) \, q_{iv}^D (x)$ and
$\delta_T \bar q_i^D (x) 
           = \alpha_{\bar q} \, \delta_T w(x) \, \bar q_i^D (x)$.
Namely, the tensor-polarized distributions are the corresponding
unpolarized ones multiplied by a weight function $\delta_T w(x)$
and a constant $\alpha_{\bar q}$. We consider that a certain fraction
of the unpolarized distributions are tensor polarized.
In expressing the unpolarized PDFs in the deuteron,
nuclear corrections are neglected,
isospin symmetry is used for the neutron PDFs
in relating them to the proton ones, and 
flavor-symmetric tensor-polarized antiquark distributions
are used in $\delta_T \bar q$.
The weight function is then parametrized 
\begin{equation}
\delta_T w(x) = a x^b (1-x)^c (x_0-x) ,
\label{eqn:dw(x)-abc}
\end{equation}
where a function with a node is taken
by considering a constraint based on a quark-parton model: 
$ \int dx \, b_1 (x)  = - \frac{5}{24} 
  \lim_{t \rightarrow 0} t F_Q (t) = 0 $ \cite{b1-sum}.
Because of this constraint, $x_0$ is expressed by the other parameters
which are determined by a $\chi^2$ analysis of the HERMES $b_1$ data.
We tried two different analyses:

\vspace{0.15cm} 
$\! \! \! \! \bullet$ Set 1: Tensor-polarized antiquark distributions 
              are terminated ($\alpha_{\bar q}$=0).

\vspace{0.15cm}
$\! \! \! \! \bullet$ Set 2: Finite tensor-polarized antiquark distributions
             are allowed ($\alpha_{\bar q}$ is a parameter).

\vspace{0.15cm}
\noindent
The parameters are determined by a $\chi^2$ fit to the data
in Fig. 10, and obtained $\chi^2$ values are $\chi^2$/d.o.f.=2.83 
and 1.57 for set-1 and 2, respectively. As shown in Fig. 10,
the set-1 analysis is not good enough to explain the HERMES data
at small $x<0.1$ without the antiquark tensor polarization.
The dashed curve is much below the data points, whereas the solid
curve can describe the data reasonably well.
From these analyses, the determined tensor-polarized distributions
for valence quarks and antiquarks are shown in Fig. 11.
Of course, the HERMES data are not accurate enough to obtain
reliable tensor distributions at this stage; however, the obtained
distributions can be used for comparing them with theoretical
calculations and for proposed future experiments such as the ones
at JLab \cite{jlab-b1}.

\vspace{0.0cm}
\begin{figure}[b]
\begin{minipage}[c]{\textwidth}
   \includegraphics[width=0.46\textwidth]{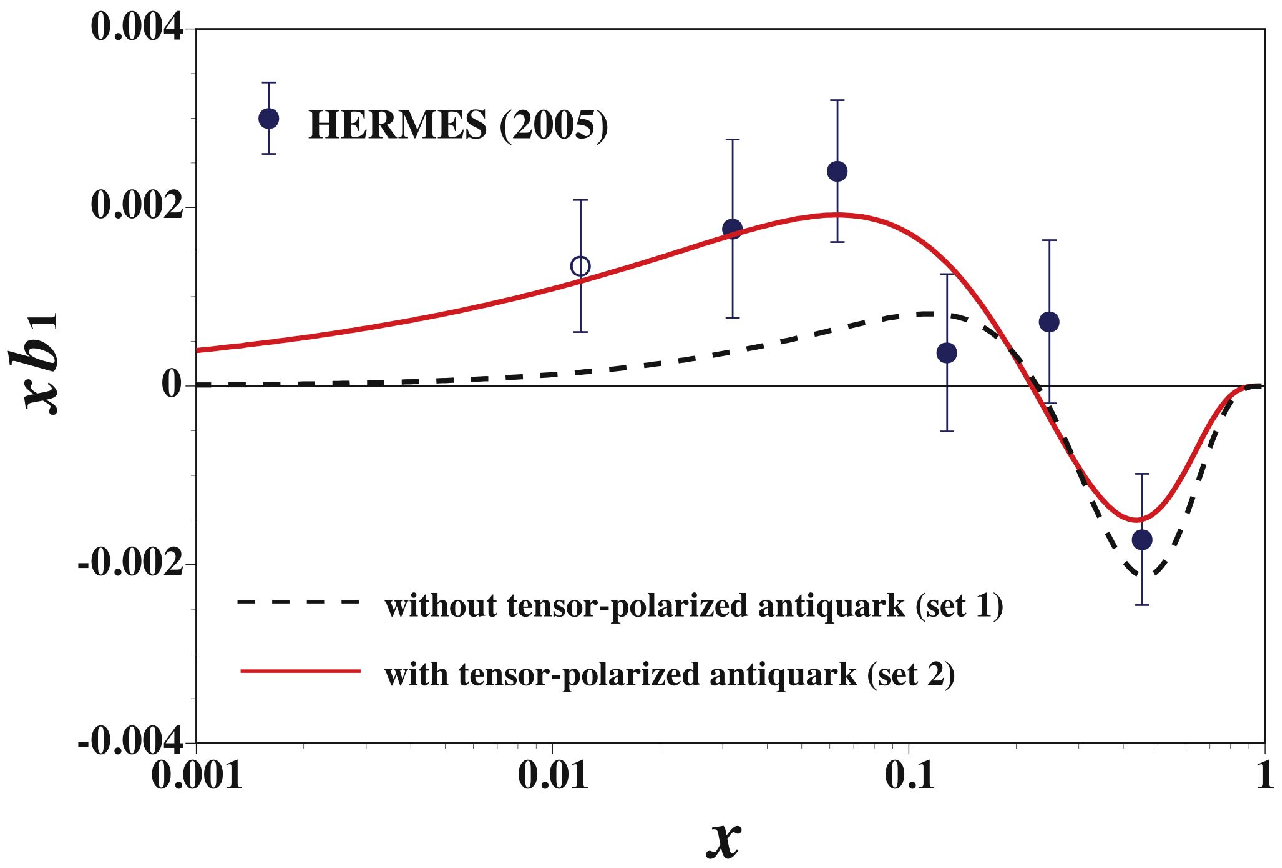}
   \hspace{0.6cm}
   \includegraphics[width=0.46\textwidth]{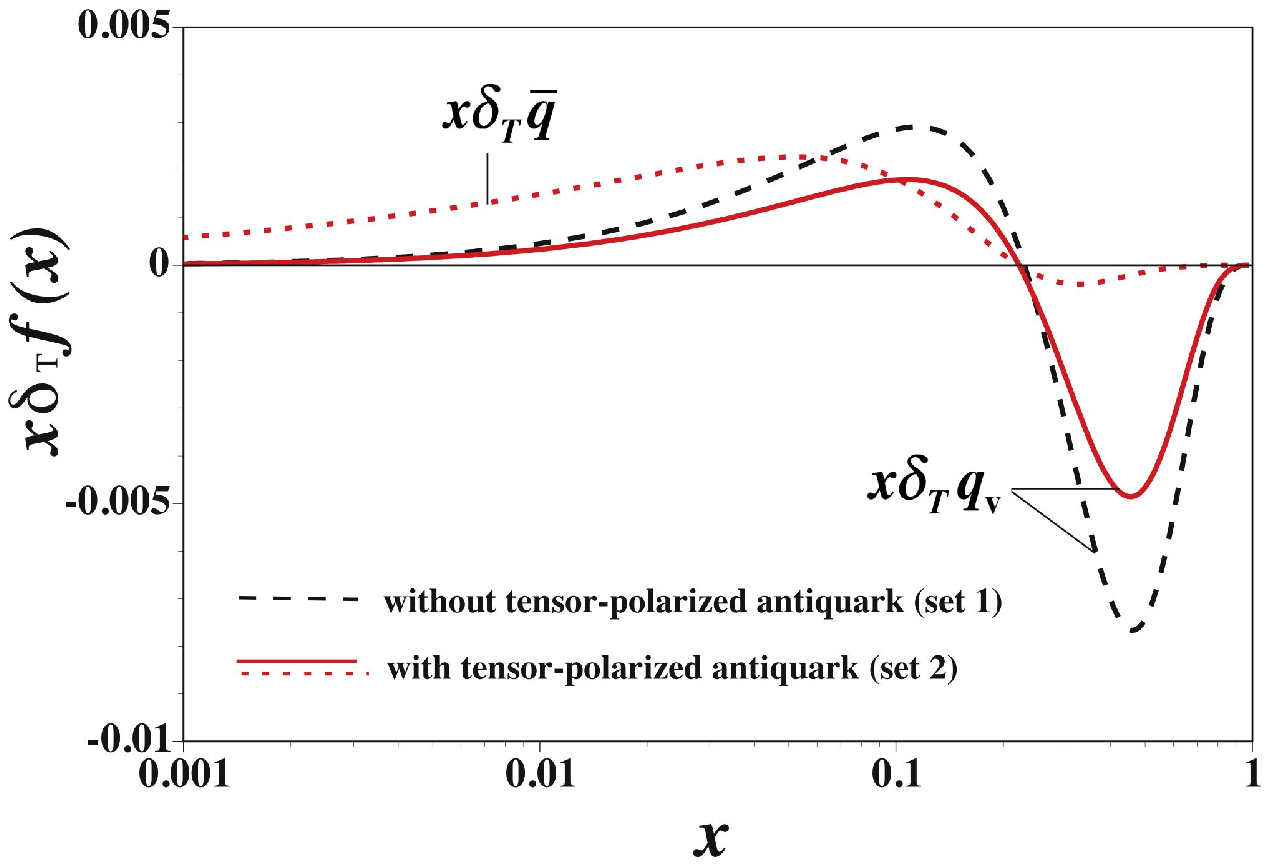}
  $ \ \ $
\label{fig:b1-motivation}
\begin{minipage}[c]{0.47\textwidth}
\setlength{\baselineskip}{10pt}
\vspace{0.3cm}
\noindent
{\footnotesize 
 {\bf FIGURE 10.} 
 Two analysis results are shown with the HERMES $b_1$ data
 \cite{sk-b1-analysis}. The solid (dashed) curve indicates
 the parametrization with (without) the antiquark tensor
 polarization.
}
\end{minipage}
\hspace{0.55cm}
\begin{minipage}[c]{0.47\textwidth}
\setlength{\baselineskip}{10pt}
\vspace{0.3cm}
{\footnotesize 
 {\bf FIGURE 11.} Obtained tensor-polarized distributions
                 \cite{sk-b1-analysis}.
        The solid and dotted (dashed) curves indicate the distributions
        with (without) the antiquark tensor polarization.
 }
\end{minipage}
\end{minipage}
\end{figure}

The $b_1$ sum is expected to vanish in a simple quark-parton model.
However, if a deviation is found, it suggests a finite tensor-polarized
antiquark distributions \cite{b1-sum}:
\begin{align}
& \int dx \, b_1 (x) = - \frac{5}{24} \lim_{t \rightarrow 0} t F_Q (t)
  + \frac{1}{18} \int dx \, [ \, 8 \delta_T \bar u (x) 
  + 2 \delta_T \bar d(x)  +\delta_T s (x) + \delta_T \bar s(x) \, ] ,
\nonumber \\
& \int \frac{dx}{x} \, [F_2^p (x) - F_2^n (x) ] = 
   \frac{1}{3} \int dx [ u_v(x) - d_v (x) ]
  +\frac{2}{3} \int dx [ \bar u(x) - \bar d(x) ] .
\label{eqn:b1x-sum}               
\end{align}
The $b_1$ sum rule is very similar to the Gottfried one \cite{flavor},
whose violation indicated a difference between $\bar u$ and $\bar d$
by the above equation. 
An interesting point of our analysis \cite{sk-b1-analysis} is that
a finite sum $\int dx b_1 (x) = 0.0058$ was obtained to suggest
a finite tensor polarization for antiquarks. It is interesting 
to conjecture a possible mechanism of this finite antiquark 
polarization and to test it by an independent experiment. 
For probing antiquark distributions, Drell-Yan process are ideal. 
In fact, polarized proton-deuteron Drell-Yan processes are
theoretically formulated in Ref. \cite{pd-drell-yan}
and they could be measured by future hadron facilities such as
the J-PARC \cite{j-parc}. The studies of the tensor structure
function $b_1$ and the tensor-polarized distributions
should shed light on a new aspect of high-energy hadron spin physics.

\section{Acknowledgements}
\vspace{-0.2cm}
The authors thank C. A. Salgado, I. Schienbein,
Journal of High Energy Physics, and American Physical Society
for permitting them to use figures from their publications.



\end{document}